\documentclass[aps,prl,twocolumn,showpacs,preprintnumbers,amsmath,amssymb]{revtex4-1}

\usepackage{graphicx}
\usepackage{dcolumn}
\usepackage{bm}
\usepackage{natbib}
\usepackage{hyperref}
\usepackage{epstopdf}

\newcommand{\expec}[1]{\langle #1 \rangle}

\newcommand{\abs}[1]{\vert #1 \vert}
\newcommand{\ket}[1]{\vert #1 \rangle}

\newcommand{\up}{\uparrow}
\newcommand{\dow}{\downarrow}

\newcommand{\h}{\hat}

\newcommand{\ie}{i.e.\ }

\newcommand{\tx}{\text}

\begin{document}

\title{Supercurrent through grain boundaries in the presence of strong correlations}

\author{F.~A.~Wolf}
\author{S.~Graser}
\author{F.~Loder}
\author{T.~Kopp}
\affiliation{
Center for Electronic Correlations and Magnetism,
Institute of Physics,
University of Augsburg,
D-86135 Augsburg,
Germany
}

\date{\today}

\begin{abstract}
Strong correlations are known to severely reduce the mobility of charge carriers near half-filling
and thus have an important influence on the current carrying properties of grain boundaries in the
high-$T_c$ cuprates. In this work we present an extension of the Gutzwiller projection approach
to treat electronic correlations below as well as above half-filling consistently. We apply this 
method to investigate the critical current through grain boundaries with a wide range of misalignment angles
for electron- and  hole-doped systems. For the latter excellent agreement with experimental data is found. We
further provide a detailed comparison to an analogous weak-coupling evaluation. 
\end{abstract}


\maketitle

\paragraph{Introduction.---} 
The manufacturing of high-temperature superconducting cables as well as the characterization
of well-defined interfaces and contacts for superconducting microwave electronics 
have been optimized over the last 20 years so that they are now found in applications
ranging from short range power supply to medical instrumentation. This was made possible by 
intensive experimental and theoretical research, which has
 led to a better understanding of the bulk as well as the
interface properties of high-$T_{\rm c}$ materials 
\cite{hilgenkamp02,dimos98,gurevich98,stolbov99,pennycook00,tanaka95,freericks06,yokoyama07,schwingschloegel09}. 
One of the most complex and puzzling problems in this respect 
is the strong reduction of the critical current $j_{\rm c}$ as a function of the grain boundary (GB) angle, 
which has been extensively studied on samples with artificially fabricated, 
well defined GBs~\cite{hilgenkamp02}. Although great progress
has been made to reduce the influence of large angle GBs on the total current~\cite{hammerl02} and to locally improve their 
current carrying properties~\cite{hammerl00},
a full theoretical understanding of the grain boundary problem is hampered by the complexity
of the physics involved. 
In a recent study the experimentally observed decay of $j_{\rm c}$ with increasing misalignment angle was reproduced in 
a microscopic model in which charge inhomogeneities at the grain boundary were identified as the main source for 
the suppression of $j_{\rm c}$~\cite{graser10}. 
But still the overall magnitude of $j_{\rm c}$ could not be correctly determined 
and it was speculated that strong electronic correlations present in the high-$T_{\rm c}$ materials might be 
responsible for this discrepancy~\cite{andersen08}. 

In this article we present an analysis of the 
dependence of the supercurrent on the grain boundary angle
in the presence of strong Coulomb interactions.
In order to describe the generic lattice defects for a 
certain GB angle, a theoretically reconstructed sample of a GB should include at least several hundred lattice sites.
Due to numerical limitations, this can only be modeled by an effective one-particle description.
To incorporate static correlations nevertheless, 
we employ a Gutzwiller approximation~\cite{zhang88} that
should already capture an important part of the GB physics. 
The Gutzwiller approach has had considerable success for 
homogeneous models of cuprate high-$T_\tx{c}$ superconductors~\cite{anderson04}.
Only in recent years was it rigorously extended
to inhomogeneous systems~\cite{wang06,ko07}, 
but is now an established method
to treat important aspects of the interplay
between impurities and superconductivity~\cite{garg08}.

\begin{figure}[tb]
\begin{center}
\includegraphics[width=0.9\columnwidth,clip]{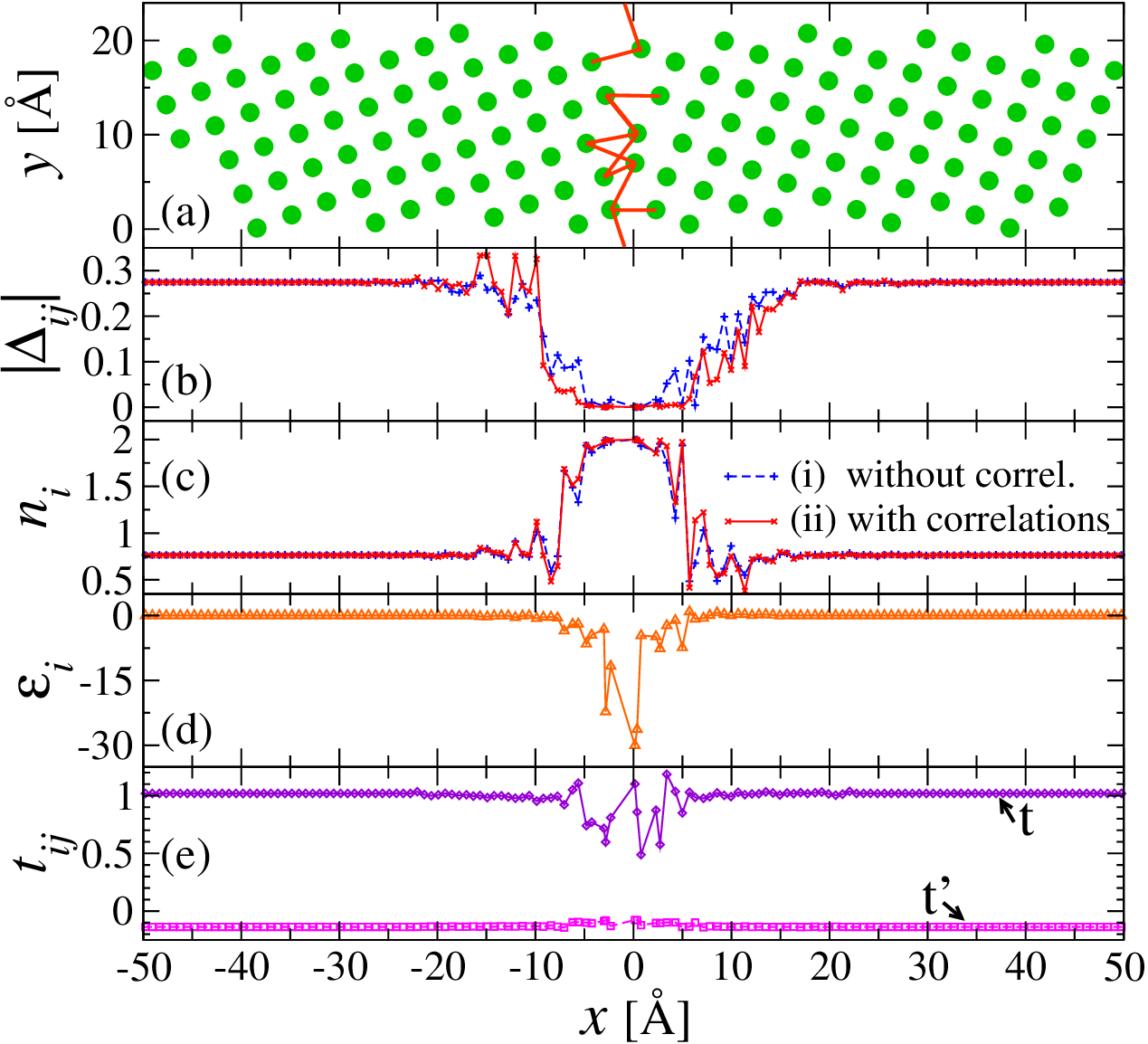}
\end{center}
\caption{The lattice structure for a (520) GB
with misalignment angle $\alpha_{(520)}=44^\circ$ 
in the $x$,$y$-plane (a). 
Orange lines mark the bonds defining the
``channels'' through the GB that explicitly appear
in the sum of Eq.~\eqref{currGB}.
Results of a self-consistent calculation for the (520) GB at a bulk filling of 
$n=0.8$ without (i) and with (ii) Gutzwiller factors 
include the $d$-wave pairing amplitude $\Delta_{ij}$ (b)
and the local density $n_i$ (c). 
In addition the local impurity potentials $\varepsilon_i$ (d)
and the averages over all nearest neighbor ($t$) and next-nearest neighbor ($t'$) 
hopping parameters of single sites~$i$ (e) are shown.
Energies are measured in units of $t_\tx{bulk}=1$.
The spin-interaction is calculated from the nearest-neighbor 
entries of $t_{ij}$ for (i) as $J_{ij} = 2.38 t_{ij}$ and for
(ii) as $J_{ij}=0.9 t_{ij}$ so that $\Delta_\tx{bulk}=0.275t_\tx{bulk}$ has the same
value for both calculations.
The precise values of the impurity potentials $\varepsilon_i$ are obtained
from the charge fluctuations assuming a screening length $l=2\tx{\AA}$.
\label{sysPar}} 
\end{figure}

The model system of a GB is, in a certain respect, a particularly inhomogeneous system
that deals with very strong impurities that to our knowledge have not yet been treated within the Gutzwiller approach.
The energy scale of the system is set by the hopping parameter $t_\text{bulk}$ in the homogeneous bulk regions far away from the GB.
In cuprates we typically expect a Coulomb repulsion of $U \sim 20\,t_\text{bulk}$
and a spin exchange interaction $J_{ij} \sim t_{ij}$, where $t_{ij}$ is the hopping matrix element for nearest and next-nearest neighbor sites.
It has been shown that charge barriers at the GB lead to local impurity potentials
of the order of $\varepsilon_i \sim -20\, t_\text{bulk}$~\cite{graser10} or more, 
therefore close to the GB, the estimate $\varepsilon_i\sim -U$ applies. 
Although in the bulk regions of a hole-doped system, the local charge densities $n_i$ 
are below half filling, the positive charges in the vicinity of the GB can induce sites with $n_i$ 
well above half filling, even in the presence of strong Coulomb repulsion.
As an example, in Fig.~\ref{sysPar}(c), (d) and (e) local charge densities, impurity potentials
and hopping amplitudes are shown as a function of the distance from a GB 
with misalignment angle $\alpha=44^\circ$. 
Such systems with strongly positively charged regions
cannot be described by the standard Gutzwiller approximation.
In the first part of this article we present an extended Gutzwiller formalism 
appropriate to describe such systems. 
In the second part we analyze our results for the critical current calculated with 
the new method and show that they are in excellent agreement with experiments.

\paragraph{Model and methodology.---}
We use the Gutzwiller projected $t\text{--}J$ model to describe the GB system. 
Within the Hartree-Fock decoupling scheme it is given by the Hamiltonian
\begin{align} \label{ham}
{\cal H} = 
& - \sum_{ijs} (g_{ij}^t t_{ij} + \chi_{ij}^*  )  c_{is}^\dagger c_{js} \nonumber\\
&-  \sum_{{ij}} (\Delta_{ij} {c_{j\up}^\dagger c_{i\dow}^\dagger} + \text{h.c.})
- \sum_{i} \mu_i \h n_{i}  \\
\text{with} \quad 
&\Delta_{ij} = 
 (\tfrac{3}{4} g_{ij}^J + \tfrac{1}{4}) J_{ij} \widetilde\Delta_{ij} ,\\
& \chi_{ij} = 
(\tfrac{3}{4} g_{ij}^J - \tfrac{1}{4}) J_{ij} \widetilde\chi_{ij},
\end{align}
where 
$\widetilde\Delta_{ij} = \frac{1}{2} (\expec{c_{i\dow}c_{j\up}} + \expec{c_{j\dow}c_{i\up}})$,
$\widetilde\chi_{ij} = \frac{1}{2}(\expec{c_{i\up}^\dagger c_{j\up}} + \expec{c_{i\dow}^\dagger c_{j\dow}})$,
$\hat n_i=\sum_sc^\dag_{is}c_{is}$,
and
$\mu_{i} = \mu - \varepsilon_i$. 
The matrix for the tunneling amplitudes $t_{ij}$ 
has non-zero entries for nearest and next-nearest 
neighbor hopping. The tunneling amplitudes $t_{ij}$  and the 
local potentials $\varepsilon_i$ are calculated using the molecular
dynamics algorithm of Ref.~\onlinecite{graser10} and
reflect the charge barriers and defects of the lattice at the GB. 
The spin-interaction coefficients $J_{ij}$ are obtained 
by rescaling the nearest-neighbor 
entries of $t_{ij}$ with a constant factor.

The Gutzwiller factors in \eqref{ham} are defined as follows:
$g_{ij}^t  = g_{i}^t g_{j}^t$ 
and $g_{ij}^J  = g_{i}^J g_{j}^J$, where 
\begin{subequations} \label{gutzfactors}
\begin{align}
g_{i}^t 
&=
\sqrt{\frac{2\abs{1-n_i}}{\abs{1-n_i} +1}},
\\
g_{i}^J 
& = 
\frac{2}{\abs{1-n_i} +1}.
\end{align}
\end{subequations}
For local densities $n_i \leq 1$ these expressions coincide with the commonly employed definitions 
of Ref.~\onlinecite{wang06}. 
By introducing the absolute values in their definitions we obtain a
renormalization that is symmetric with respect to half filling $n_i=1$.
This is equivalent to a local particle-hole transformation for sites where $n_i>1$.
Due to the particle-hole invariance of the Hubbard model from which 
the $t$--$J$ model is derived, this treatment is obviously
correct for a homogeneous system with $n_i= n > 1$. 

The rigorous derivation for an inhomogeneous system is
based on the definition of a projection operator $\mathcal{P}=\prod_i \mathcal{P}_i$ 
that locally projects either doubly occupied or empty
sites out of a BCS type wave function $\ket{\psi}_0$, depending on the local density:
\begin{equation} \label{project}
\mathcal{P}_i
=
\left\{
\begin{array}{ll}
y_i^{\hat n_i} (1-D_i) & \text{if } n_i \leq 1 \\
y_i^{\hat n_i} (1-E_i) & \text{if } n_i > 1 \,.
\end{array}
\right. 
\end{equation}
Here, $y_i$ is a fugacity factor, while 
$D_i= \h n_{i\up} \h n_{i\dow}$ and $E_i=(1- \h n_{i\up}) (1- \h n_{i\dow})$ 
denote the projection on doubly occupied and empty sites, respectively.  
Note that the selection of different projectors depending on the local density 
is a natural extension of the operator $y_i^{\h n_i}$:
this operator assigns different weights
to contributions from empty and singly occupied sites and thus ensures 
a projected state $\ket{\psi}$ that can be transparently
compared with the pre-projected state $\ket{\psi_0}$ due to its
common set of local densities~\cite{wang06}.
A comparison of expectation values evaluated with the wave functions 
$\ket{\psi_0}$ and $\ket{\psi}= \mathcal{P} \ket{\psi_0}$, 
completely analogous to the one presented in Ref.~\onlinecite{wang06},
yields the results of Eq.~\eqref{gutzfactors}.

It remains to be justified that the implementation of a symmetric
renormalization around half filling, or equivalently, projecting out empty and doubly occupied sites in the 
free wave function, is physically meaningful. 
The argumentation here is very similar to the original idea of Gutzwiller, to project doublons out of the 
wave function due to their energy cost. 
Our extension is obviously justified for bulk regions: 
In an environment where $\mu_i$ is large enough to yield $n_i>1$, a local particle-hole 
transformation
leads to the removal of empty sites from the wave function. 
Also at the boundary of such a region the definition of Eq.~\eqref{project} 
is consistent since on a site with $\mu_i \sim U$, the gain of potential 
energy cancels the cost of double occupancy, independent of the situation on neighboring sites. 
Thus the energy cost to create a holon at site $i$ is 
of the size of $U$ and consequently it is projected out.

Note that this procedure is possible in 
the Gutzwiller approximation only due to its fundamental assumption
that inter-site correlations can be neglected. 
However, in the framework of the Gutzwiller approach it is a 
consistent improvement of the wave function
modeling in the spirit of assigning different weights to different
sites with the fugacity factor $y_i^{\h n_i}$.
It is a heuristic, very detailed and sophisticated method of 
minimizing the energy of a system described in
a phenomenological mean-field picture. Below 
we will show that this picture is sufficiently realistic to accurately describe
the physics of superconducting charge transport through a GB.

The renormalized model Hamiltonian of Eq.~\eqref{ham} is evaluated 
within the Bogoliubov--de Gennes (BdG) framework.
We enforce constant phases with a difference $\phi=\pi/2$ on the
pairing amplitude $\Delta_{ij}$ 
on either end of the sample, in the center of which the GB is situated. 
The corresponding phase gradient
induces finite currents given by
\begin{equation} \label{currBond}
j_{ij} = - 2 g_{ij}^t t_{ij} \sum_s \text{Im}(\expec{c_{is}^\dagger c_{js}}).
\end{equation}
The normalized total current through the GB is
\begin{equation} \label{currGB}
j_\tx{c} = \frac{\sin(\alpha/2)}{M_y} \sum_{x_i <0, x_j>0} j_{ij},
\end{equation}
where $x_i$ and $x_j$ are the $x$-coordinates of sites on either side of the GB,
$\alpha$ is the misalignment angle and $M_y$ the second Miller index of the $(M_x M_y M_z)$ GB.
The factor ${\sin(\alpha/2)}/{M_y}$ normalizes the current to a unit length in $y$-direction.
To be able to compare with experimental results we employ
the following relation \cite{graser10} between the current $j_\tx{c}^{\tx{abs}}$
in absolute units ${\tx{A}}/{\tx{cm}^2}$ and the current in units $[t_{\tx{bulk}}]=\tx{eV}$
\begin{equation}
j_\tx{c}^{\tx{abs}} 
= 
\frac{\Delta^\tx{exp}}{\Delta^\tx{theo}} 
\frac{N_\tx{UC}}{ac} \frac{e}{\hbar} \, j_{\rm{c}} \, ,
\end{equation}
where $N_\tx{UC}=2$ is the number of CuO$_2$ planes per unit cell, 
$\alpha$ the misalignment angle and
$a=3.82\tx{\AA}$ and $c=11.7\tx{\AA}$ the lattice spacing
in $x$- and $z$-direction, respectively.
The values
$\Delta^\tx{exp}\sim0.025 \tx{eV}$
and
$\Delta^\tx{theo}=\Delta_\tx{bulk}=0.275 \tx{eV}$
are valid  for the bulk regions.
In order to compensate the smaller magnitude of the experimental quantities we employ the factor
$\Delta^\tx{exp}/\Delta^\tx{theo}\simeq 0.09$
where we make use of the fact that $j_\tx{c}$ scales
approximately linearly w.r.t.\ $\Delta$ if $\Delta$ is small. 
We cannot employ the experimental value $\Delta^\tx{exp}\sim0.025 \tx{eV}$
in our calculations as for gap values of this order of magnitude,
the iterative solution of the BdG equations is non-convergent.

\begin{figure}[!tb]
\begin{center}
\includegraphics[width=0.9\columnwidth,clip]{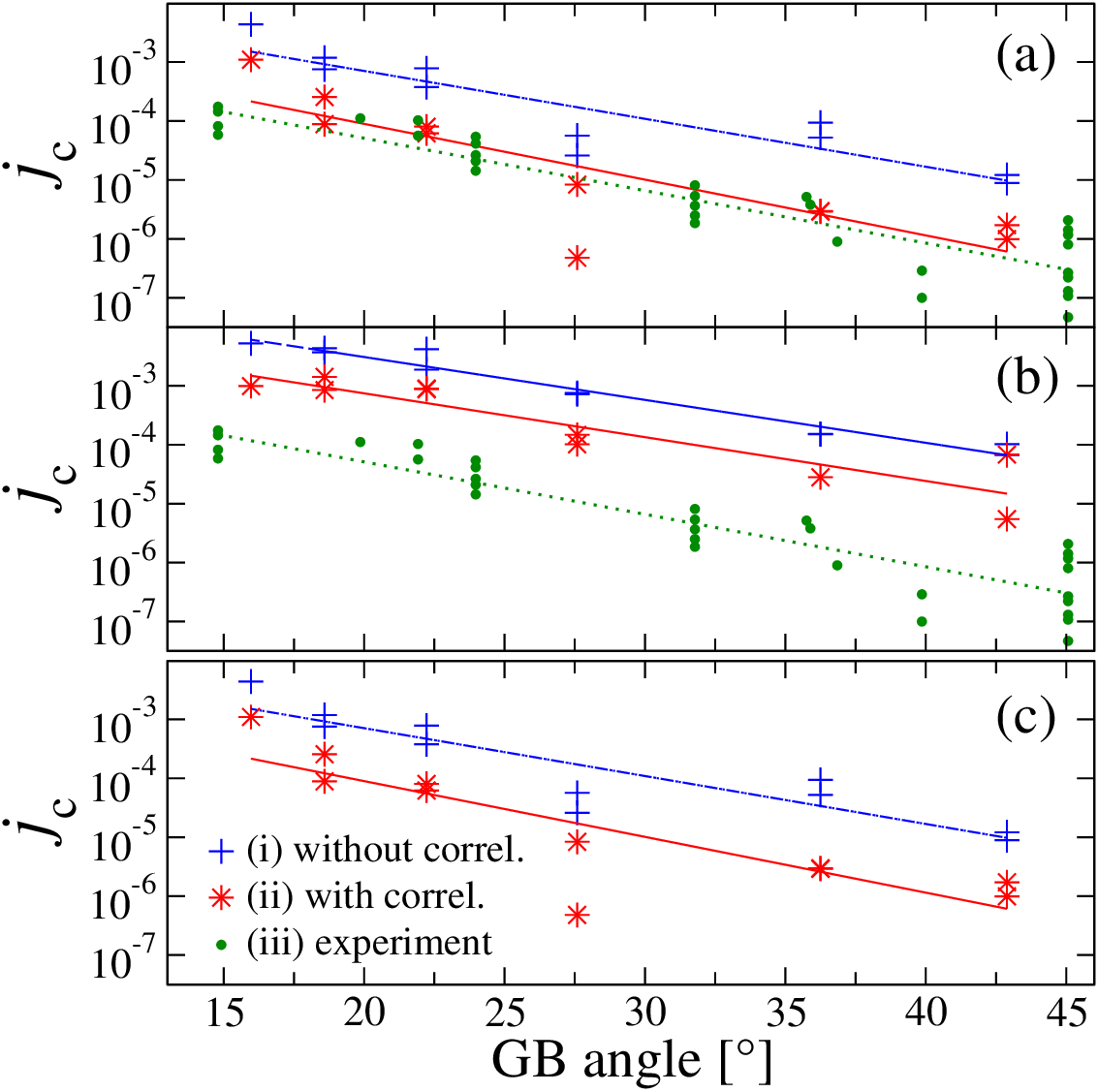}
\end{center}
\caption{Angular dependence of the critical current for
GB samples with electron density $n=0.8$ (a,b) and $n=1.2$ (c). We plot results
for 11 samples for 6 different angles in between $15^\circ$ and $45^\circ$ in order
to be able to average. This average is calculated as a linear
fit to the logarithm of the data points and depicted as colored lines.
The impurity potentials have been
calculated from the charge fluctuations assuming 
a screening length of $l=2 \tx{\AA}$ (a) and $l=1.2\tx{\AA}$ (b,c).
The current is given in units of $t_\tx{bulk}$.
Depicted are calculations using
the BdG Hamiltonian of Eq.~\eqref{ham} where for (i)
all Gutzwiller factors are set to $1$ and
for (ii) all Gutzwiller factors are taken as defined in Eq.~\eqref{gutzfactors}.
The theoretical values in (a) and (b) are compared to
experimental data (iii) for optimally doped YBCO thin films on SrTiO$_3$
taken from Ref.~\onlinecite{hilgenkamp02}.
The spin-interaction is calculated from the nearest-neighbor 
entries of $t_{ij}$ for (i) as $J_{ij} = 2.38 t_{ij}$ and for
(ii) as $J_{ij}=0.9 t_{ij}$ so that 
$\Delta_\tx{bulk}=0.275t_\tx{bulk}$ has the same
value for both calculations.
\label{figCurrAngle}} 
\end{figure}

\paragraph{Angle dependence of the supercurrent.---}
In Fig.~\ref{figCurrAngle}(a) and (b) we show on a logarithmic scale 
the dependence of the zero-temperature supercurrent on the GB misalignment angle 
for a hole-doped system. 
We compare two types of calculations with experimental
data taken from Ref.~\onlinecite{hilgenkamp02}:
(i) a standard evaluation of the GB-Hamiltonian within the BdG framework 
(\ie $g_{ij}^t = g_{ij}^J = 1$) and 
(ii) an evaluation with the local Gutzwiller factors as defined in Eq.~\eqref{gutzfactors}.
In order to provide a meaningful comparison of (i) and (ii)
we employ the same hopping matrices $t_{ij}$
and impurity potentials $\varepsilon_i$ for all calculations, 
but globally scale $J_{ij}$ to obtain identical values for the $d$-wave 
pairing amplitude in the bulk: $\Delta_\text{bulk}=0.275\,t_\text{bulk}$.
This approach is appropriate as $j_\tx{c} \propto \Delta_\text{bulk}$ and it is
in complete analogy to Ref.~\onlinecite{garg08}.
Note that also at the GB the pairing amplitude $\Delta_{ij}$ 
shown in Fig.~\ref{sysPar}(a)
does not differ qualitatively for the evaluation 
without (i) and with (ii) Gutzwiller factors.

As can be seen from Fig.~\ref{figCurrAngle} the supercurrent decays 
in both cases (i) and (ii) exponentially with the GB angle.
We also find that the current in the correlated system (ii) is almost
one order of magnitude smaller
than in the system without correlations (i).
As the screening length $l$ in the CuO$_2$ planes is only approximately known we performed
two calculations assuming different values of $l$, with $l=2\tx{\AA}$ for Fig.~\ref{figCurrAngle}(a)  and $l=1.2\tx{\AA}$
for Fig.~\ref{figCurrAngle}(b). For the choice of $l=2\tx{\AA}$ we obtain excellent agreement
with the experimentally determined critical currents over a wide range of GB angles. 

In Fig.~\ref{figCurrAngle}(c) we show the angle dependence of the current for an electron-doped system.
Comparing panels (b) and (c) of Fig.~\ref{figCurrAngle},
we find that the current is almost particle-hole symmetric.
Correlation effects are known to be strongest
at half filling where a Mott insulator is obtained in the Gutzwiller approximation. 
The dominant reduction of the current due to correlations
should therefore be caused by areas in the system where $n_i\sim1$.
These areas are equally present in the hole- and electron-doped system
in the transition region to the bulk.
In the hole-doped case such a region is already implied by the
fact that, directly at the GB, the system is always well above half filling.
In the electron-doped case it is caused by large density fluctuations. These
appear due to the charge fluctuations with the same magnitude 
in the electron- and the hole-doped case.
For a hole-doped example, Fig.~\ref{sysPar}(c)
shows values for the density with $0.5\leq n_i \leq2$.
Obviously, these arguments imply that the correlation
effects depend on the charge fluctuations. 
Furthermore, the fact, that the narrow transition regions have the same extension for all angles, 
explains that the exponential behavior 
w.r.t.\ the GB angle appears in both calculations (i) and (ii):
only the break-up of pairs by i	mpurities, present in (i) and (ii), 
causes the exponential behavior while the correlation induced reduction
of the supercurrent is constant for all angles.

\begin{figure}
\begin{center}
\includegraphics[width=\columnwidth,clip]{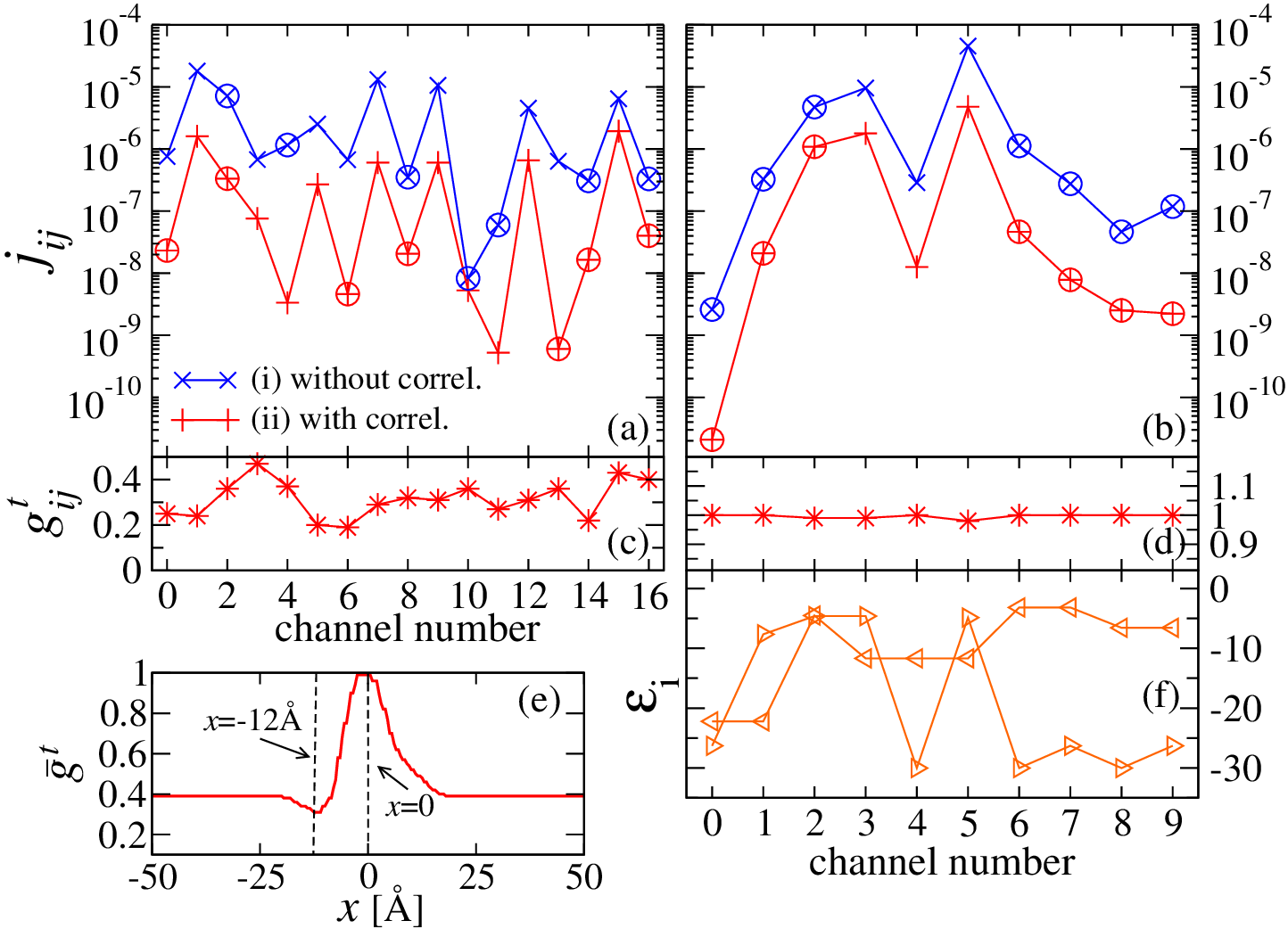}
\end{center}
\caption{Cuts parallel to a (520) GB with angle $\alpha_{(520)} \sim 44^\circ$ 
in $y$-direction at $x= -12${\AA} (a,c) and $x=0$ (b,d,f).
In the first row (a,b) we show the values for the currents  $j_{ij}$ through all channels of a cut.
The channels of the $x=0$ cut are depicted in Fig.~\ref{sysPar}(a) as orange lines.
Negative currents are marked with a circle around the data point.
In the second row (c,d) we show the Gutzwiller factors $g_{ij}^t$ and in (f) 
the local impurity potentials $\varepsilon_{i}$ of the two sites that are connected by a channel.
In panel (e) the average $\bar{g}^t$ of $g_{ij}^t$ along cuts for fixed values of $x$ is displayed.
\label{figChannels}} 
\end{figure}

Fig.~\ref{figChannels} displays the microscopic (local) 
values of the currents $j_{ij}$, 
the Gutzwiller factors $g_{ij}^t$ and the local potentials $\varepsilon_{i}$
for cuts parallel to the GB. Panels (a,c) show these values at the 
edge of the GB (at $x= -12${\AA}) where potential fluctuations 
are minor but where the average $\bar g^t$ of $g_{ij}^t$ along the cut 
is minimal, indicating the strongest
suppression of hopping by Coulomb repulsion (see panel (e)). 
Panels (b,d,f) refer to a cut along the center of the GB (at $x= 0$), 
where the potential fluctuations are maximal but the Gutzwiller factors 
$g_{ij}^t$ are close to 1 due to the strong effective electron doping
induced by the local impurity potentials.
The flat distribution of $g^t_{ij}$ in panel (d) 
leads to the homogeneous reduction for all channels in panel (b)
while in panel (c) $g^t_{ij}$ varies significantly which results in a
non-homogeneous reduction of $j_{ij}$ in panel (a).
Current conservation guarantees that the total 
reduction of the supercurrent in (b) equals
the total reduction in (a).
Both panels (a,b) show that charge transport 
is carried almost entirely by very few channels.
At the GB center these channels are exposed only to slightly 
varying impurity potentials in contrast to the ones that
carry almost no current (panel (f)).
Even at the edge of the grain boundary, where almost 
no potential fluctuations are present, 
back currents continue to exist (panel (a)). They are induced by the 
impurities at the center of the GB and degrade slowly
on account of the small phase gradient away from the GB.
Finally we point out that fluctuations of the 
critical current are strong in experiment and that they
are better reproduced by the correlated 
than by the non-correlated modeling  (see Fig.~\ref{figCurrAngle}(a,b)).
Random Coulomb blocking of  
some of the few channels that carry almost all current produces these
correlation induced fluctuations.

\paragraph{Conclusion.---}
We expect that the extension of the Gutzwiller projection
approach, which we introduced to treat strongly inhomogeneous systems,
is not only viable for the characterization of the
supercurrent through grain boundaries. It should also be
instrumental for other systems with intrinsic impurities 
or artificial structures (heterostructures).
We emphasize that we achieved convincing agreement
with experiment for the supercurrent through GBs.
The reduction of the current was shown to be due to 
the intertwining effects of large charge fluctuations
and strong correlations. It can thus be understood
that experimental doping of the GB in order to moderate 
these fluctuations significantly improves
the current carrying properties, which has already been observed
for Ca-doped interfaces \cite{hammerl02,hammerl00}.

\begin{acknowledgments}
The authors gratefully acknowledge discussions with
C.~W.~Schneider, B.~M.~Andersen, P.~J.~Hirschfeld, and J.~Mannhart.
This work was supported by the DFG through TRR80.
\end{acknowledgments}

\end{document}